\documentclass[final,5p,times,twocolumn]{elsarticle}
\usepackage{subfigure}

\usepackage{graphicx}
\usepackage{multirow}
\usepackage{amssymb}

\journal{journal for publication}

\begin{document}

\begin{frontmatter}


\author{Elisa Londero\corref{cor2}}
\ead{londero@chalmers}
\author{Elsebeth Schr{\"o}der}
\cortext[cor2]{Corresponding author}
\address{Microtechnology and Nanoscience, MC2, Chalmers University of Technology, SE-412 96 G{\"o}teborg, Sweden}

\title{Vanadium pentoxide (V$_{2}$O$_{5}$): a van der Waals density functional study}
\hyphenation{wide-spread}
\hyphenation{cor-re-la-tion}



\begin{abstract}
The past few years has brought renewed focus on the physics behind the class of materials characterized by long-range interactions and wide regions of low electron density, sparse matter.
There is now much work on developing the appropriate algorithms and codes able to correctly describe this class of materials within a parameter-free quantum physical description. 
In particular, van der Waals (vdW) forces play a major role in building up material cohesion in sparse matter. 
This work presents an application to the vanadium pentoxide (V$_2$O$_5$) bulk structure of two versions of the vdW-DF method, a first-principles procedure for the inclusion of vdW interactions in the context of density functional theory (DFT). 
In addition to showing improvement compared to traditional semilocal calculations of DFT, we discuss the choice of various exchange functionals and point out issues that may arise when treating systems with  large amounts of vacuum.  
\end{abstract}


\end{frontmatter}


\section{Introduction}
\label{1}
Density functional theory (DFT) is one of the most reliable and widespread methods for atomic-level computational studies of the structure and electronic properties of materials. 
The formalism behind DFT is exact, but for practical use an approximation for the exchange-correlation part $E_{xc}$ of the total energy functional must be found. 

Dense crystalline materials show two main characteristics: they have a periodic structure and they are characterized by strong bonds. 
The periodic structure  simplifies the description in codes using periodic boundary conditions, and the strong bonds are well described already by one of the simplest of the approximations for $E_{xc}$, the local-density approximation (LDA), and also by the later, semilocal, generalized gradient approximation (GGA).
For this reason dense crystalline materials have been studied systematically and in great detail for a long time. 
On the other hand the extension to account for matter with regions of low electron density and a considerable component of long-ranged forces, like the van der Waals (vdW) forces, has proven to be more difficult.
The vdW interaction is a quantum-mechanical phenomenon providing bonding by correlating instantaneous charge fluctuations.  
A successful method that is able to describe the vdW forces, the vdW-DF method, has been developed relatively recently \cite{Dion,Thonhauser,vdW-DF2}.

\begin{figure}[bth]
\begin{center}
\includegraphics[width=0.45\textwidth]{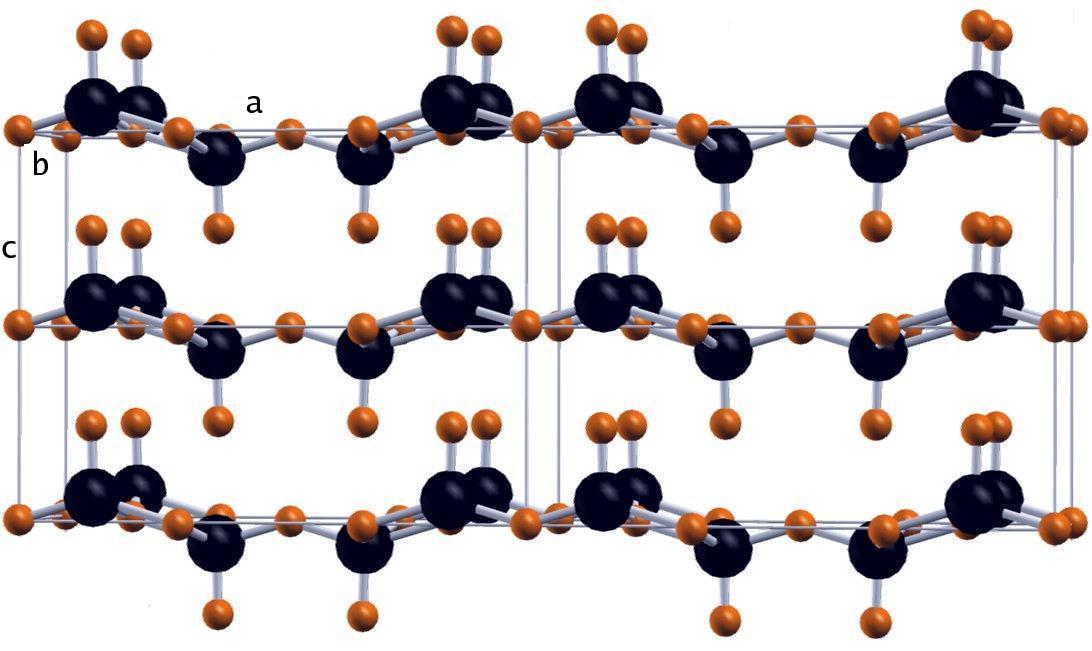}
\caption{Illustration of the V$_2$O$_5$ bulk structure.
The unit cell is here repeated four times to illustrate the layered structure.
The lattice parameters $a$, $b$ and $c$ are indicated.
The black spheres are the vanadium atoms and the small spheres are the oxygen atoms \cite{xcrysden}.}
\label{fig:1}
\end{center}
\end{figure}

In this paper we present a DFT study using the vdW-DF method for vanadium pentoxide (V$_2$O$_5$). 
The $\alpha$-phase of V$_2$O$_5$ has a layered structure, illustrated in Figure \ref{fig:1}.
The bonds within the layers are strong while the interlayer binding is long-ranged and weaker and for this reason the material is easy to cleave in the plane perpendicular to $c$.

V$_2$O$_5$ has previously been studied with DFT by a number of groups using GGA \cite{Pirovano,Reeswinkel,Goclon,Xiao,hejduk,sauer2008} or by adding \mbox{(semi-)empirical} vdW-terms to GGA calculations \cite{sauer2008}.
Here our focus is on the vdW-DF method rather than on the material V$_2$O$_5$ itself. We therefore mainly discuss the issues arising when using vdW-DF in a layered, extended system such as V$_2$O$_5$ bulk. 

The structure of this paper is as follows. 
In Section 2 we explain the computational methods used, focusing in particular on the technicalities required both by the self-consistent GGA calculations and by the vdW-DF postprocess procedure that allows the inclusion of vdW forces. 
In Section 3 we show and discuss our results, and Section 4 contains a summary.  

\section{Calculational methods}
\label{2}
Vanadium pentoxide has a layered structure with charge voids between the layers, and for this reason a DFT analysis requires use of the vdW-DF method. 
The calculation consists in moving the vanadium pentoxide layers apart in the stacking direction, accordion-like. 
We study the changes in total energy with changes in the values of $c$.
The intralayer binding, with a mixed covalent and ionic nature, is a strong binding and this justifies our choice of keeping the interatomic distances fixed inside the layer itself when the layers are taken apart.

The computational procedure is made up by two steps. 
First a self-consistent GGA calculation using the plane-wave DFT code \textsc{dacapo} \cite{dacapo} is carried out and subsequently the vdW-DF method is applied to calculate the nonlocal energies in a post-GGA procedure. 
This has been shown to not produce significant discrepancies in the energy calculations, compared to an entirely self-consistent vdW-DF calculation \cite{Thonhauser}.

Focusing on the details of the self-consistent GGA calculations, the Brillouin zone is sampled using a  2$\times$4$\times$4 Monkhorst-Pack $k$-point set and both the plane-wave and density cutoffs are set to 500 eV. 
We check the convergence of our calculations with respect to these parameters. 
The orthorhombic unit cell contains two V$_2$O$_5$ formula units and it is periodically repeated in the three spatial directions.
We use in-plane lattice constants $a=11.55$ {\AA} and $b=3.58$ {\AA}, which are the optimal values in GGA calculations.  
Ultrasoft pseudopotentials (USPP) are used. 
The fast Fourier transform (FFT) grid is chosen such as to have at most 0.12 {\AA} between nearest-neighbor grid points. 
For each FFT grid point the electron density $n(\mathbf{r})$ is self-consistently calculated within GGA and subsequently a vdW-DF calculation is performed starting from $n(\mathbf{r})$. 

We take isolated layers of V$_2$O$_5$ as the reference point of the energy.  
The reference calculations are carried out using a unit cell that has a length in the $c$ direction that is 4 times the length of the bulk unit cell, with a single layer of V$_2$O$_5$ in the middle of it, surrounded by vacuum.

For the inclusion of the vdW interactions in our calculations we use the vdW-DF scheme implemented in a non-periodic code. 
The correlation energy $E_c$ is divided into two parts, one part that is mostly local and approximated by the LDA correlation energy $E_c^{\mathrm{LDA}}$, and one part that includes the most nonlocal interactions and is written \cite{Dion}:
\begin{equation}\label{eq:1}
E_c^{\mathrm{nl}}[n]=
\frac{1}{2}\int\int d\mathbf{r}\,d\mathbf{r}'\,n(\mathbf{r})\phi(\mathbf{r},\mathbf{r}')n(\mathbf{r}') \,.
\end{equation}
The correlation term is thus
\begin{equation}
E_c=E_c^{\mathrm{LDA}}+E_c^{\mathrm{nl}}
\end{equation}
and it is added to the total energy from the self-consistent GGA calculation after 
removing the GGA correlation $E_c^{\mathrm{GGA}}$.

As mentioned above, the code used for this second step is not periodic in space so the spacial periodicity of the material is reproduced by adding a number of cells in the three directions of space (for the bulk calculation).
More precisely, for each electron density grid point $\mathbf{r}$ within the central unit cell, all the other grid points $\mathbf{r}'$ to be used within the integral (\ref{eq:1}) fall in two regions, limited by two radii.
Inside the smallest radius all available grid points are used in the $E_c^{\mathrm{nl}}$ calculation. In the other region---inside the larger radius but outside the smaller radius---only half of the grid points in any of the three spacial directions are used \cite{berland}.
The reference calculations (isolated layers) are carried out similarly, with the exception that no additional grid points are included in the stacking direction $c$.  
The two radii are chosen such that further inclusion of grid points only changes the energy contribution from  $E_c^{\mathrm{nl}}$ marginally. 

At this point another computational choice must be taken: the most appropriate exchange functional to be used. 
The vdW attraction is purely a correlation effect so binding from exchange that mimics the vdW binding must be avoided. 
The revPBE exchange functional has often been used because for some examples of sparse matter it has been shown to give the least spurious exchange contribution $E_x$ to the binding energy \cite{Dion}.
However, revPBE exchange is overly repulsive, and it is relevant to also consider other choices of exchange functionals. 
The exchange forms used here are discussed in Section 3.     

There are two relevant versions of the vdW-DF method presently available.\footnote{A preceding version \cite{rydberg,layerPRL} was developed for layered material in which the layers are approximately translationally invariant. This is not the case for  V$_2$O$_5$ and thus that version is ignored here.} 
The first \cite{Dion,Thonhauser}, which is here called vdW-DF1, has shown to work well by improving binding energies and separation distances over the GGA results. 
It has been tested \cite{langrethjpcm2009} on a variety of systems among those: atomic and molecular systems (such as dimers of benzene, benzene-like molecules, and polycyclic aromatic hydrocarbons (PAH), and polymer interactions), crystalline solids (like graphite, potassium intercalation in graphite), and adsorption (benzene and adenine on graphite, PAHs on graphite, adsorption of aromatic and conjugated compounds on MoS$_2$). 
Despite these successes, vdW-DF1 underestimates the hydrogen-bond strength and overestimates equilibrium separations.

The slightly different version of the vdW-DF method, called vdW-DF2 \cite{vdW-DF2}, has recently been developed. 
It has been constructed focusing on obtaining more accurate energies for finite, relatively small molecules, but here we test it for the extended system vanadium pentoxide. 
It uses a different expression of the plasmon frequencies used in the evaluation of $E_c^{\mathrm{nl}}$. 
To reach a description within chemical accuracy for most of the molecules of the S22 set, the use of a different semilocal exchange functional is also suggested in Ref.\ \cite{vdW-DF2}.   
The exchange choice suggested in Ref.\ \cite{vdW-DF2} to go with vdW-DF2 is the refitted version of PW86 \cite{PW86,Murray}.

In the present paper we combine the two versions of vdW-DF with a number of exchange choices (subscript $x$ denotes the exchange part of the functional): 
the revPBE$_x$ \cite{revPBE} and PW86$_x$ choices mentioned above, and the recently suggested PBEsol$_x$ \cite{PBEsol}, optB88$_x$ \cite{optB88}, and C09$_x$ \cite{cooper}.

\section{Results and discussion}
\label{3}

We have previously presented V$_2$O$_5$ bulk binding results \cite{vanadium} found by use of vdW-DF1.  
The choice of the exchange functional, even if not the major step of the whole procedure, is of importance and the selection or the construction of the most appropriate exchange form is still a matter of debate. 

In this spirit we here use both vdW-DF1 and vdW-DF2 in combination with five different forms of exchange, some of which for physical reasons may be considered good candidates.

\begin{figure}[h]
\begin{center}
\includegraphics[width=0.45\textwidth]{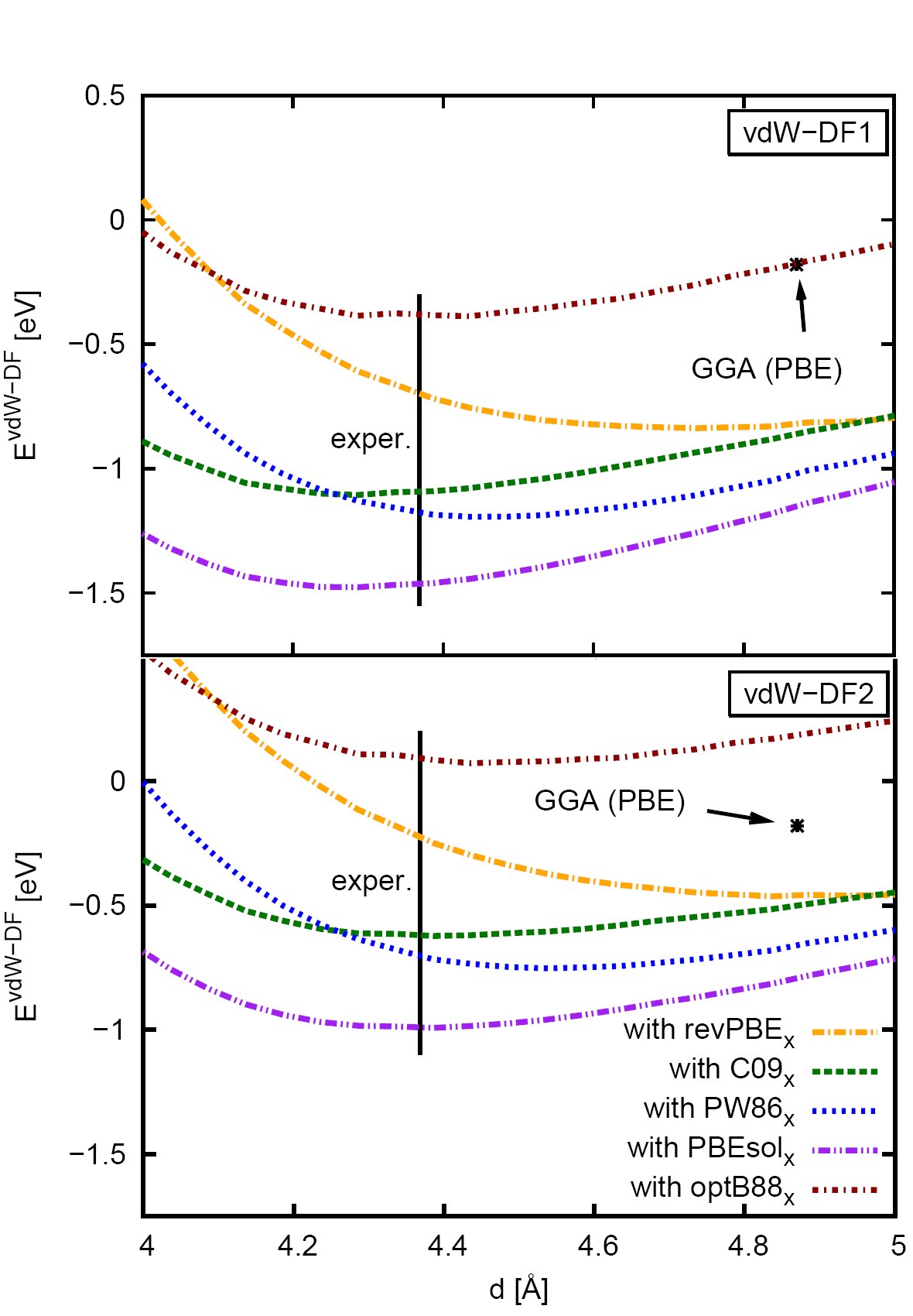}
\caption{Top panel: vdW-DF1 energy curves for different choices of the exchange functional compared with the results for vdW-DF2 (bottom panel). 
The vertical black lines show the optimal binding distance (no binding energy available) according to experiments \cite{Enjalbert}. 
The stars show the position of the minimum according to DFT calculations (GGA) that do not account for vdW interactions.} 
\label{fig:2}
\end{center}
\end{figure}

\subsection{Interlayer binding results}
Figure \ref{fig:2} shows the resulting total energy curves, $E^{\mathrm{vdW-DF}}(c)$, obtained from bulk calculations with the separated oxide layers as reference calculations.
As described in detail elsewhere \cite{vanadium,PAHgg} we ensure that the atomic positions on the underlying FFT grid in the reference calculation are identical to those of the bulk calculation.  
As discussed further below, we have excluded the contribution to $E_c^{\mathrm{nl}}$ from a few grid points in the vacuum region that have unphysical values of $n$.
The minima of the $E^{\mathrm{vdW-DF}}(c)$ curves (the binding energies $E_b$ with positive values for binding systems) are summarized in Table \ref{tab:1}, along with experimental results and a GGA calculation, using PBE \cite{PBE}, without inclusion of vdW forces. 

\begin{table}[h]
\begin{center}
\begin{tabular}{llcc}
\hline\hline
&                        &$c$ [\AA]      & $E_b$ [eV]\\
\hline
\textit{vdW-DF1}  
&\textbf{revPBE$_x$}     & \textbf{4.72} & \textbf{0.86} \\
&PW86$_x$                & 4.46          & 1.19   \\
&C09$_x$                 & 4.28          & 1.10   \\
&PBEsol$_x$              & 4.28          & 1.48   \\
&optB88$_x$              & 4.37          & 0.39   \\
\hline
\textit{vdW-DF2} 
&revPBE$_x$              & 4.87          & 0.48   \\
&\textbf{PW86$_x$}       & \textbf{4.55} & \textbf{0.75}   \\
&C09$_x$                 & 4.38          & 0.62   \\
&PBEsol$_x$              & 4.36          & 0.99   \\
&optB88$_x$              & 4.47          & $-$0.08  \\ 
\hline  
      \textit{Comparison} 
&PBE                     & 4.87          & 0.18  \\
&Experiment$^{\mathrm{a}}$& 4.368        & -     \\
      \hline\hline
\multicolumn{4}{l}{$^{\mathrm{a}}$Ref.\ \cite{Enjalbert}.} 
    \end{tabular}
\caption{\label{tab:1} Results for the equilibrium distances and the binding energies per unit cell. 
The results are shown for both vdW-DF1 and vdW-DF2 with various exchange functionals.
The boldface entries are the defaults for the vdW-DF version.}
\end{center}
\end{table}

\begin{figure}[h]
\begin{center}
\includegraphics[width=0.45\textwidth]{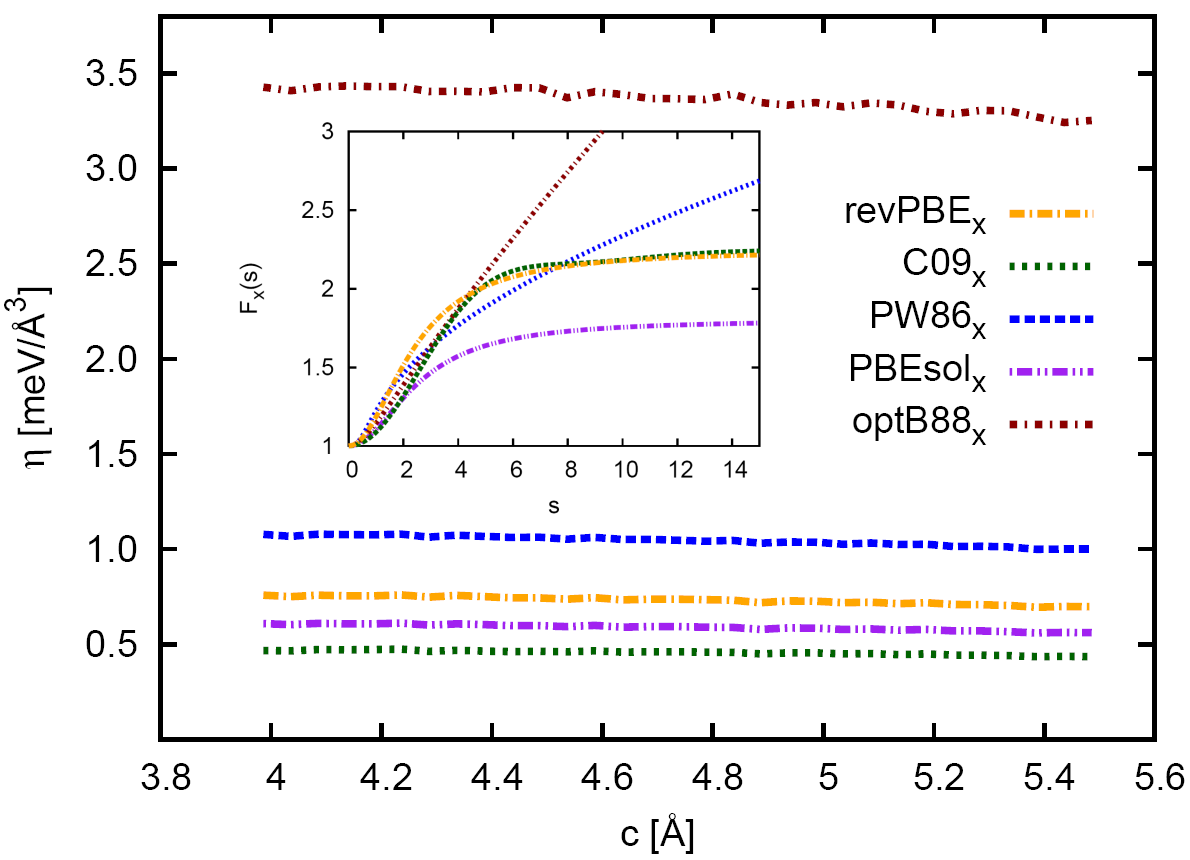}
\caption{Error measure $\eta$ for the contribution of pseudopotential-related ``noise" to the exchange energy if not explicitly excluded.
Insert: Enhancement factor $F_x(s)$ for the exchange energy.}
\label{fig:noise}
\end{center}
\end{figure}

The various exchange choices of GGA type that we consider here differ by the choice of enhancement factor $F_x(s)$ in the exchange energy density.
The exchange energy may be written 
\begin{equation}
E_x = \int d\mathbf{r} \, n(\mathbf{r})\, \epsilon_x^{\mathrm{LDA}}\left(n(\mathbf{r})\right)\, F_x\left(s(\mathbf{r})\right)
\label{eq:fxint}
\end{equation}
with $\epsilon_x^{\mathrm{LDA}}\sim n^{1/3}$ and where $s=c_s |\nabla n|/n^{4/3}$ with $c_s=1/(2\pi^{2/3} 3^{1/3})$ is the reduced electron density gradient.
We plot $F_x(s)$ for various exchange forms in the insert of Figure \ref{fig:noise}.

Several physical constraints may be imposed on the form of $F_x(s)$, as described in the literature \cite{PBEcomment}. 
In the limit of small $s$ (approximately $0 < s < 1$), the functionals PBEsol$_x$ and C09$_x$ are designed to follow the  simple gradient expansion approximation (GEA) form.
The small-$s$ region corresponds to the region of large densities $n$.
This was suggested in Ref.\ \cite{PBEsol} to remove an artificial bias towards free atoms and in Ref.\ \cite{cooper} to reduce the short-range repulsion compared to revPBE$_x$.
The other three exchange choices (revPBE$_x$, PW86$_x$, and optB88$_x$) all have steeper growth of $F_x(s)$ for small $s$ (Figure \ref{fig:noise}).

In the large-$s$ limit it has been argued \cite{Harris,Murray} that constant asymptotes of $F_x(s)$ may lead to the same type of spurious binding from exchange alone as in LDA.
This arises in regions that have both low values of the density $n$ and an inhomogeneous distribution (large $\nabla n$), for example outside surfaces and in internal voids.
In Ref.\ \cite{Murray} it is argued that for small molecules a growth of $\sim s^{2/5}$ gives the best agreement with Hartree-Fock results.
However, in our present system with extended oxide layers this is not necessarily the case.
Of the exchange forms used here the revPBE$_x$, C09$_x$, and PBEsol$_x$ tend to constants at large $s$, PW86$_x$ grows as $\sim s^{2/5}$, and optB88$_x$ has a much steeper growth ($\sim s$).

Figure \ref{fig:2} and Table  \ref{tab:1} show that in general, the improvement of the equilibrium distances using the vdW-DF technique is evident when compared to a standard GGA calculation.
It is also clear that the suppression of $F_x(s)$ for small $s$ in PBEsol$_x$ and C09$_x$ results in smaller binding distances. 
Of all exchange choices presented here, PBEsol$_x$ and C09$_x$ give the smallest binding separations. 
For vdW-DF2 those separations are very close to the experimental value, for vdW-DF1 they are a few percent smaller than the experimental value.

\subsection{Numerical noise}

Soft matter have substantial regions of low electron density.
In such systems the electron density can be as low as $10^{-9}$ $|e|/${\AA}$^3$.
An issue related to the presence of such densities is that the reduced density gradient $s\sim\nabla n/n^{4/3}$ that enters the enhancement factor $F_{x}(s)$ becomes extremely large unless the small density is locally almost constant ($\nabla n$ small).
Depending on the shape of $F_{x}(s)$, the exchange energy (\ref{eq:fxint}) can be more or less influenced by those large $s$ values.

Our use of USPP and, more generally, noise in planewave code calculations, gives rise to an additional problem: the presence of small unphysical negative electron density values.
The \textsc{dacapo} code formally tries to solves this problem by introducing a floor for the electron density, $n_{\mbox{\scriptsize floor, dac}} = 10^{-9}$ $|e|/${\AA}$^3$, and setting all the negative values equal to the floor before evaluating the $E_x$ term.
This procedure of course increases the number of very small but positive $n$ values that give rise to very large values of $s$.

\begin{figure}[h]
\begin{center}
\includegraphics[width=0.45\textwidth]{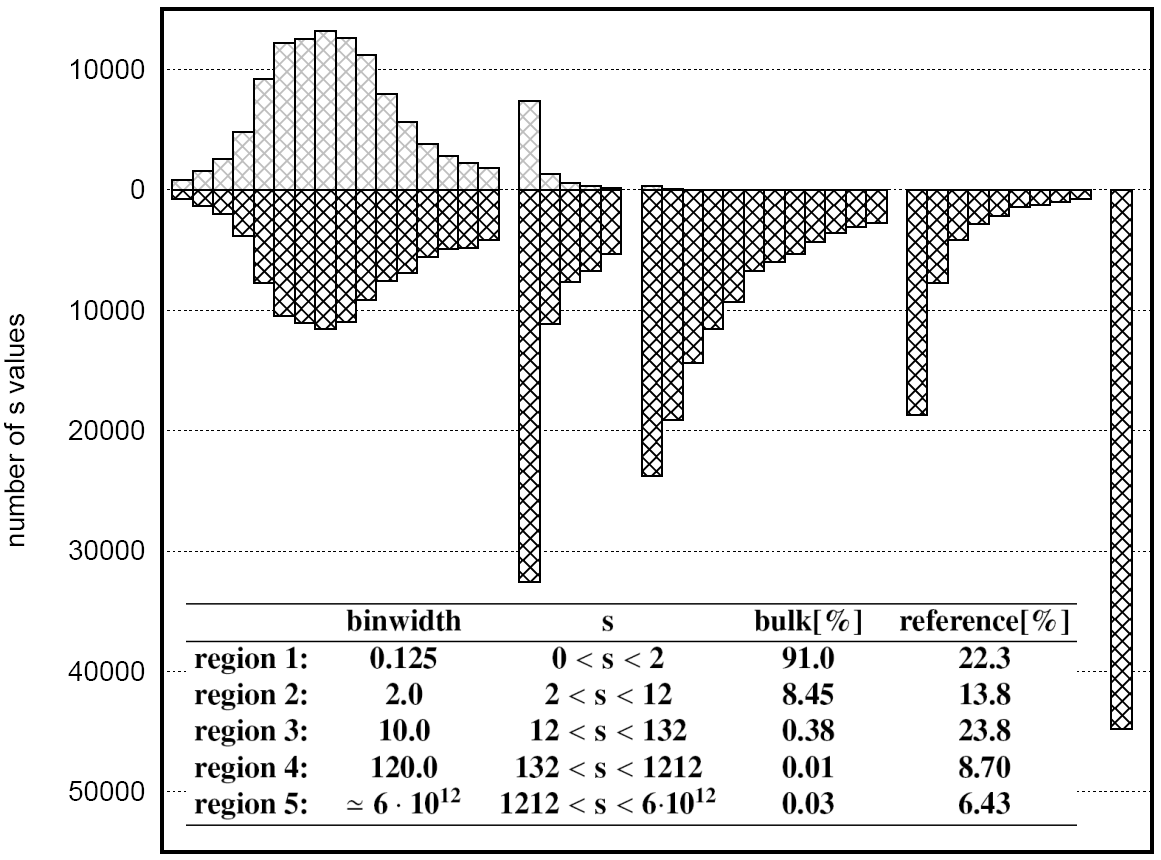}
\caption{Distribution of $s$ values for V$_2$O$_5$ at the equilibrium distance. 
Values are obtained from dacapo generated GGA based electron charge densities $n$. 
Top part: bulk. Bottom part: reference calculation (isolated V$_2$O$_5$ layer with much vacuum). 
The corresponding distribution for bulk of dense materials is usually in the range $0 <s< 3$.}
\label{fig:histo}
\end{center}
\end{figure}

In Figure \ref{fig:histo} we show the distribution of $s$ values for both a bulk and a reference calculation of the V$_2$O$_5$ structure. 
The plot is based on the valence electron density $n$ from the \textsc{dacapo} GGA calculations. 
For evaluating $s$ for the histogram in Figure \ref{fig:histo} we have introduced the same type of replacement of negative $n$ with a small positive value, since negative $n$ make no sense in the evaluation of $s$. 
We here use the value $n_{\mbox{\scriptsize floor}}=10^{-15}$ $|e|/${\AA}$^3$, but the distribution is not sensitive to the precise value of the $n_{\mbox{\scriptsize floor}}$.  
In the bulk calculation hardly any negative values of $n$ appear ($\sim 30$ grid points have negative values out of a total of $\sim 10^{5}$ grid points), and in the reference calculation all values in the region-5 bin ($s > 1212$) originate in point that had $n<n_{\mbox{\scriptsize floor}}$ before the replacement. 

The small values of $n$, leading to relatively large values of $s$ in \textsc{dacapo} $E_x$ calculations, are thus a mix of physically correct small values and ``noise" arising from the USPP description of $n$.
In order to limit contributions to $E_x$ from the low-density values (most of them unphysical), we carry out our own $E_x$ calculations, based on the dacapo valence density $n$ but now with \textit{removal\/} of all contributions from grid points with $n < n_{\mbox{\scriptsize floor}} = 10^{-15}$ $|e|/${\AA}$^3$.
To illustrate the effect of this removal we have calculated the quantity:
\begin{equation}
\eta=\frac{1}{V}\, \Delta\!\!\! \int_{n<n_\mathrm{floor}} \!\!\!\!\!\!
d\mathbf{r}\, |n| \, \epsilon_x^{\mathrm{LDA}}\left(|n|\right)\, F_x\left(c_s \frac{|\nabla n|}{|n|^{4/3}}\right);
\end{equation}
which is the difference (as expressed by $\Delta$) between the integral of the exchange energy for the bulk structure and the integral for the reference structure.
The \textbf{r} spans the density grid points that have a value of $n$ below the floor $n_{\mbox{\scriptsize floor}}$, i.e., the points we remove in our calculations. 
The error measure $\eta$ includes a division by the volume of the (bulk) unit cell $V = abc$ because more (physical) vacuum area enters as the layers are moved apart, and this physical contribution is not to be seen as an error contribution.
The results are shown in the main panel of Figure \ref{fig:noise}.

We find that for each choice of exchange the measure $\eta$ is approximately constant.
This is in agreement with our expectation that contributions to $\eta$ come mainly from the vacuum part of the unit cell in the reference calculation, a region that grows approximately linearly with unit cell volume $V$ and thus $c$.
We find that for most of the exchange functionals the error measure is small, at approximately 0.4 to 1.1 meV/{\AA}$^3$, but for optB88$_x$ it is much larger, at 3.4 meV/{\AA}$^3$.
This means that unless we explicitly exclude those points, the optB88$_x$ will contribute to the $E_x$ an erroneous 0.6 eV per unit cell at the binding distance.

Noise on top of larger electron densities still affects our calculations.
This is so even if we have, by the exclusion, removed an important part of the erroneous contribution.
The error measure should therefore only be taken as an indication of the sensitivity of the exchange choice to possible numerical noise in $n$.
It is clear that optB88$_x$ is much more sensitive that the other choices, to the extent that we will not include optB88$_x$ in our work on vdW-DF development.
Nevertheless, it is interesting to analyse the behavior of optB88$_x$ to learn about the numerical challenges which exist for accurate calculations with the vdW-DF method. 

Why is the optB88$_x$ so much more sensitive to noise than the other exchange choices we consider here? 
It is clear from the form of optB88$_x$ that numerical noise in systems with large regions of low densities $n$ will have large contributions to $E_x$ from these regions.
The growth of optB88$_x$ for large $s$ is so steep, $F_x\sim s \sim n^{-4/3}$ (even if $|\nabla n|$ is assumed constant at a finite value), that the factor $n \epsilon_x^{\mathrm{LDA}}\sim n^{4/3}$ in the integrand will not suffice to damp the contribution. We have
\begin{eqnarray}
E_{x,\mathrm{voids}}^{\mathrm{optB88}}&=& 
\int_{\mathrm{voids}} d\mathbf{r} \, n(\mathbf{r})\, \epsilon_x^{\mathrm{LDA}}\left(n(\mathbf{r})\right)\, F_x^{\mathrm{optB88}}\left(s(\mathbf{r})\right)
\\
&\approx&
 \mathrm{constant} \int_{\mathrm{voids}}  d\mathbf{r} \, |\nabla n(\mathbf{r})|
\end{eqnarray}
and the latter expression is obviously very sensitive to noise.
The $F_x$ for the other exchange choices grow less steep, or not at all, at large $s$, and the contribution from the void areas is thus damped by $n \epsilon_x^{\mathrm{LDA}}$.

While the USPP is in our case likely the main origin of noise (of which we remove an important part explicitly), this is a problem not only for \textsc{dacapo} or other codes that use USPP. 
Most DFT codes were written at a time when small electron densities were not relevant and the codes are thus not optimized for handling regions of very low electron densities.

We do not want to give recommendations of any particular exchange form based on the one system (V$_2$O$_5$) studied here.
However, we do show that calculations are sensitive to the choice.
We also find that optB88$_x$ is not a valid candidate in systems like the present both for physical reasons ($F_x(s)$ growth too steep as discussed in Ref.\ \cite{Murray}) as well as for numerical reasons as discussed above.

The effect of the noise on the binding energies found in Table \ref{tab:1} is severe for the case of optB88$_x$ (estimated error 0.6 eV), but smaller for the other exchange choices (0.1 to 0.2 eV). 
It is important to note that these estimates are approximate, that they may include also physically correct small values of $n$ (although this is not the case in this particular system), and that noise on the larger values of $n$ is not taken into account.

\section{Summary}
\label{4}
The structural properties of a vanadium pentoxide bulk structure are calculated within the vdW-DF method, testing both of the presently available version: vdW-DF1 of Ref.\ \cite{Dion} and vdW-DF2 of Ref.\ \cite{vdW-DF2}. 
A number of forms of the exchange functional are tested.                   
The interplanar distance, which is usually severely overestimated in traditional GGA calculations, is found to be closer to the experimental value using these methods.
This confirms the expectation that the interplanar bond in vanadium pentoxide has a large vdW component.

\section{Acknowledgments}
We thank M.V.\ Ganduglia-Pirovano for discussions about V$_2$O$_5$, and J.\ Rohrer, K.\ Berland, and P.\ Hyldgaard for discussions of exchange functional issues.
Partial support from the Swedish Research Council (VR) is gratefully acknowledged, as well as allocation of computer time at UNICC/C3SE (Chalmers) and SNIC (Swedish National Infra\-structure for Computing).





\bibliographystyle{model1b-num-names}
\bibliography{<your-bib-database>}

\begin{thebibliography}{99}
\bibitem{PBEcomment}
A good starting point is Ref. \cite{PBE} and its references.
\end{thebibliography}



\end{document}